%% file: acat2000serv.tex
\documentclass{aipproc}

\usepackage{epsfig}

\input{defs}

\begin{document}

\title{
\hfill {\normalsize {\rm CERN-TH/2001-253}}\\[-.3em]
\hfill {\normalsize {\rm DCPT/01/78, IPPP/01/39}}\\[-.3em]
\hfill       \mbox{\normalsize {\rm hep-ph/0109237}}\\[1em]
Feynman-diagram evaluation in the\\
electroweak theory with computer algebra}
\author{G.~Weiglein}
\affiliation{CERN, TH Division, CH--1211 Geneva 23, Switzerland}

\begin{abstract}
The evaluation of quantum corrections in the theory of the electroweak
and strong interactions via higher-order Feynman diagrams requires
complicated and laborious calculations, which however can be structured
in a strictly algorithmic way. These calculations are ideally suited for
the application of computer algebra systems, and computer algebra has
proven to be a very valuable tool in this field already over several
decades. It is sketched how computer algebra is presently applied
in evaluating the predictions of the electroweak theory with high precision, 
and some recent results obtained in this way are summarized.
\end{abstract}

\maketitle

\section{Introduction}

The electroweak and strong 
interactions of elementary particles are very successfully
described by quantized gauge field theories. 
The quantized nature of these
theories manifests itself via corrections beyond the lowest order
in the perturbative expansion, which is based on Feynman diagrams. 
The evaluation of higher-order
Feynman diagrams (which are called loop diagrams) is a technically very
complicated but on the other hand algorithmic procedure. The development
of computer-algebra systems was boosted by the demand for this kind of
applications in particle physics, and \textsl{Schoonschip}~\cite{Velt67}
was one of the first implementations of a powerful computer-algebra
program. Further examples of computer-algebra systems that have their
roots in particle physics are \textsl{Reduce}~\cite{reduce}, 
\textsl{Macsyma}~\cite{macs}, \mma~\cite{math}, and \FO~\cite{form}.

Computer-algebra systems allow to perform symbolic manipulations and
algebraic calculations without round-off errors. They are equipped with
a number of built-in algorithms which provide the basis for 
the user to implement his/her own algorithms for handling specific
problems. Present-day computer-algebra systems furthermore possess
capabilities for communicating with external programs, e.g.\ with
routines for numerical evaluation, text processing and graphics, or 
with other computer-algebra programs.

Examples of computer-algebra systems being widely used at present in
high-energy theory are \mma~\cite{math}, \map~\cite{maple} and
\FO~\cite{form}. \mma\ and \map\ are general-purpose programs containing
a large number of built-in functions (and many additional software 
packages are available). These programs offer capabilities for both symbolic 
and numerical computations, support graphical display, and possess
user-friendly interactive platforms. The application of these systems to
problems in high-energy physics involving expressions with a huge number
of terms can be limited by the computing speed or by memory problems. The
latter applies in particular to non-local operations, like e.g.\
factorization, which require to have all terms of an expression available
within the physical memory of the computer. \FO, on the other hand, is a
program that was specifically optimized for handling very large
expressions. It is less user-friendly than \mma\ and \map, containing
much fewer built-in operations and allowing only non-interactive
execution. For recent developments concerning the parallelization of \FO,
see \citere{formparall}.

\section{Perturbative evaluation of gauge theories with 
computer-algebraic methods}

The concept of treating interactions as a perturbation to a free
field theory and performing an expansion in the coupling constants leads
to a description of scattering processes in terms of Feynman diagrams.  
The lowest-order prediction, corresponding to the classical limit, for 
a process with a certain number of external particles (e.g.\ a $2 \to 2$
scattering process) is obtained from the sum of the connected diagrams
containing the lowest possible power of the coupling constants (which
enter via the interaction vertices). These are in general tree-level 
diagrams.
In higher-order diagrams additional interaction vertices give rise to
closed loops of propagators, 
for which an integration over the internal momenta has to be performed. 

The prediction for a scattering process of certain fields, assigned to
the external legs, and a specified number of loops
can be obtained via an algorithmic procedure. In a
first step, all topologically different diagrams (for which in
renormalizable theories only 3-point and 4-point interaction vertices
are possible) have to be generated. Inserting the fields of the model
under consideration into the topologies in all possible ways leads to
the Feynman diagrams. The Feynman rules translate these graphical
representations into mathematical expressions. 

Since the loop integrals in general lead to divergences, the expressions
need to be regularized (i.e.\ made mathematically meaningful). In a
renormalizable theory the divergences can be absorbed into a
redefinition of the parameters of the theory. The renormalization is
furthermore necessary in order to fix the physical meaning of the
parameters order by order. 

The evaluation of the Feynman amplitudes involves a treatment of the 
Lorentz structure of the amplitude, calculation of Dirac traces etc. 
At the one-loop level it is possible to reduce all tensor integrals 
to a set of standard scalar integrals, 
which can be expressed in terms of known analytic functions. 
As a consequence, with the existing techniques a wide class of processes with
up to four external legs can be evaluated at the one-loop order in
massive gauge theories (for a discussion of the technical problems 
occurring in one-loop processes with six external legs, see e.g.\
\citere{vicini}).

In contrast to the one-loop case, no general algorithm exists so far for
the evaluation of two-loop corrections in the electroweak theory.
The main obstacle in two-loop calculations in massive gauge theories
is the complicated structure of the two-loop integrals, which makes both
the tensor integral reduction and the evaluation of scalar integrals
very difficult. In general the occurring integrals are not expressible
in terms of polylogarithmic functions. For the evaluation of some types
of integrals that do not permit an analytic solution numerical
methods and expansions in their kinematical variables have been
developed.

Applying the appropriate on-shell conditions to the external legs one
obtains the S-matrix element from the sum of all contributing Feynman
amplitudes. Squaring it and performing the phase space integrations one
finally arrives at predictions for cross sections and life times.

Computer-algebraic methods can facilitate most of the above-mentioned
steps. Besides benefits from automation, a computer-algebraic treatment
is also useful for verifying the correctness of the different steps of a
certain calculation. In particular, results obtained at 
the algebraic level (before inserting specific numerical values for the
parameters) are well suited for highly non-trivial checks, e.g.\ with
respect to their UV- and IR-finiteness, gauge-parameter independence,
and the validity of Slavnov--Taylor identities. As an example, in
\citere{tensred} a Slavnov--Taylor for the two-loop Z-boson self-energy
in the electroweak Standard Model (SM) has been verified by showing that 
the results of about 4000 Feynman diagrams add up to zero algebraically. 

As indicated by the above example, powerful computer-algebraic tools are 
very useful for calculations (in particular of higher-order corrections)
in the SM, since the large number of different fields in the SM 
gives rise to a large number of contributing Feynman diagrams (at the
one-loop level typically $\ord(10^2)$, at the two-loop level
$\ord(10^3)$), and the massiveness of the fields makes the
evaluation of the loop diagrams very complicated in general. The
technical complications are even higher in extensions of the SM.
In the Minimal Supersymmetric Standard Model (MSSM) the duplication of the
number of fields compared to the SM leads to a plethora of possible 
interaction vertices and consequently to a large increase in the number 
of diagrams contributing at a certain order. In QCD, on the other hand, 
computer-algebraic tools are particularly valuable for multi-loop
applications. In \citere{qcdbeta}, for instance, the four-loop
$\beta$~function of QCD has been calculated. This required the computation of
about 50000 diagrams, showing clearly the need for a high degree of
automation. Similarly, for tree-level processes with many particles in the 
final state thousands of diagrams can contribute and algebraic methods can be
useful for obtaining compact and numerically efficient representations,
see e.g.~\citere{abtohl}.

Examples of computer-algebra based collections of program packages 
presently used for higher-order calculations in the electroweak theory 
and QCD are (where the different
programs in each collection mostly use common syntax and can be linked
together)
\begin{itemize}
\item[(i)]
\fa~\cite{KuBD91}, \fec~\cite{Mert91}, \foc~\cite{formcloopt}, 
\two~\cite{tensred,two},
\textsl{LoopTools}~\cite{formcloopt}, \textsl{s2lse}~\cite{s2lse},

\item[(ii)]
\textsl{GEFICOM}~\cite{harlstein}, \textsl{QGRAF}~\cite{qgr},
\textsl{MATAD}~\cite{mat}, \textsl{MINCER}~\cite{minc},

\item[(iii)]
\textsl{DIANA}~\cite{diana}, \textsl{QGRAF}~\cite{qgr}, 
\textsl{ON-SHELL2}~\cite{sh2},

\item[(iv)]
\xlo~\cite{xlo97}, \textsl{GiNaC}~\cite{ginac}.
\end{itemize} 

\fa, \fec, \foc\ and \two\ are written in \mma\ (\foc\ is partially
written in \FO). \fa\ is a program for
generating all Feynman amplitudes contributing to a certain process to a
given order in \mma\ format and for drawing the corresponding Feynman
diagrams. As a feature of particular importance for higher-order
calculations in the electroweak theory, \fa\ generates not only the
unrenormalized diagrams at a given order but also the counterterm
contributions at this order and the counterterm diagrams needed for the
subloop renormalization. The model files for the electroweak SM (including
the Feynman rules for the background-field formulation of the SM~\cite{bfm})
and QCD are predefined in \fa. Recently also the model file for the MSSM
has been completed~\cite{MSSMmodfile}.
In applications to other models, e.g.\ chiral perturbation
theory~\cite{Buerg}, the appropriate model file has to be provided by
the user. This was also the case in previous applications in the 
two Higgs-doublet model~\cite{twoHD} and the
MSSM~\cite{mssm2lqcd,mh2fd,Arhr,svenacat}.

\fec\ and \foc\ are programs (using the \fa\ syntax) for algebraically 
evaluating one-loop diagrams in the electroweak theory and QCD
with up to four external legs in a highly automatized way.
\foc\ internally uses an interface to \FO,
which is used for the memory- and time-intensive parts of the
calculation. \foc\ can directly be linked to \textsl{LoopTools}, which
contains routines for the numerical evaluation of scalar one-loop
integrals and one-loop tensor coefficients. \textsl{LoopTools} is based
on the \textsl{FF}~\cite{FF} package and provides a \textsl{Fortran} and a 
\textsl{C++} library.

As mentioned above, much less tools are available for two-loop
calculations in massive gauge theories compared to the one-loop case.
The program \two\ is based on an algorithm for the tensor reduction of 
general two-loop 2-point functions, which extends the algorithm for the
tensor reduction of one-loop integrals~\cite{pass}. \two\ can be used 
for an automatic reduction of Feynman
amplitudes for two-loop self-energies with arbitrary masses, external
momenta, and gauge parameters to a set of standard scalar integrals.
It can directly be linked to the program \textsl{s2lse}, which is 
written in \textsl{C++} and performs the evaluation of the scalar 
two-loop 2-point integrals by means of one-dimensional integral
representations in terms of elementary functions, which allow a fast
numerical evaluation with high precision.

The program collection \textsl{GEFICOM}, \textsl{QGRAF}, \textsl{MATAD},
and \textsl{MINCER} is mainly used for calculations in QCD and for
the evaluation of QCD corrections to electroweak observables. 
\textsl{GEFICOM} acts as the master program that calls the other
packages. It contains \mma\ and \textsl{Fortran} routines as well as 
elements written in the script languages \textsl{AWK} and \textsl{PERL}.
For details of \textsl{GEFICOM} and some examples of its applications, 
see \citere{harlstein}.

The \textsl{Fortran} program \textsl{QGRAF} is an efficient generator
for Feynman diagrams. As output the diagrams are encoded in a symbolic 
notation. Being optimized for high speed, \textsl{QGRAF} is particularly
useful for applications involving a very large number (i.e.\ $\ord(10^4)$) 
of diagrams. Within \textsl{GEFICOM}, the evaluation of the diagrams
proceeds by performing expansions in their kinematical variables. 
The resulting integrals are then computed with \textsl{MINCER} and
\textsl{MATAD}. The program \textsl{MINCER} performs the computation of
integrals up to three-loop order where all lines are massless and only
one external momentum is non-zero. It makes in particular use of
integration-by-parts methods~\cite{intbypart}. While its original
version was written in \textsl{Schoonschip}, the present version 
of \textsl{MINCER} is realised in \FO. The \FO\ program \textsl{MATAD}
was designed for the computation of vacuum integrals up to three-loop
order which contain only one mass scale (i.e.\ their propagators are either 
massless or carry a common mass).

The \textsl{C} program \textsl{DIANA} is designed as a master program
for higher-order calculations, i.e.\ it calls the necessary subprograms 
for a specific computation. It
reads the output of \textsl{QGRAF} and can produce a graphical 
representation for the diagrams if the relevant topologies are
pre-defined by the user. For the calculation of the diagrams \FO\
programs are called, e.g.\ the package \textsl{ON-SHELL2} which can be
used for the calculation of single-scale two-loop 2-point functions 
(diagrams with only one non-zero mass in the internal lines and 
the external momentum on the same mass shell).

\xlo\ is a \map\ package for calculating certain one-loop and two-loop 
diagrams in the electroweak SM, which is linked to \textsl{C++} routines
for numerical integration of loop integrals. The symbolic part of \xlo\
is planned to be based in the future on \textsl{GiNaC}, which is a 
specifically designed framework written in \textsl{C++}.

\section{Examples of higher-order results in the SM and the MSSM}

In the following some examples are sketched of recent higher-order 
results obtained with \fa, \two\ and \textsl{s2lse} in the electroweak
SM and the MSSM. Within the SM, higher-order calculations are necessary
for the comparison of the theory predictions with the experimental results 
for electroweak precision observables like $\MW$, $\sweff$ etc.\ which have
meanwhile reached an accuracy of better than $1 \times
10^{-3}$~\cite{summer01data}. The precision tests of the SM allow in
particular to set constraints on the mass of the Higgs boson, which is
the last missing ingredient of the SM and plays a crucial role for a
consistent description of massive particles.

\begin{figure}[htb]
\centerline{
\psfig{figure=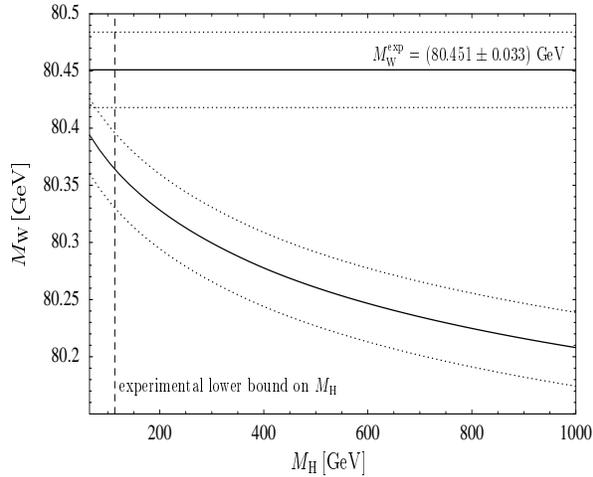,width=8cm,height=6.3cm}
}
\caption{The prediction for $\MW$ as a function of $\MH$ for
$\mt = 174.3 \pm 5.1$~GeV is compared with the current experimental
value, $\MW^{\mathrm{exp}} = 80.451 \pm 0.033$~GeV~\cite{summer01data}, and
the experimental 95\% C.L.\ lower bound on the Higgs-boson mass,
$\MH = 114.1$~GeV~\cite{LEPHiggsSM0701}.
\label{fig:mw2l}
}
\end{figure}

In \citere{delr} the currently most accurate prediction for the W-boson
mass, $\MW$, within the SM has been obtained. It contains it particular 
the complete fermionic contributions at the two-loop level, which are
treated exactly, i.e.\ without an expansion in the top-quark or the
Higgs-boson mass. The result for $\MW$ is shown in \reffi{fig:mw2l} as a
function of the Higgs-boson mass, $\MH$. It is compared with the current
experimental value for $\MW$~\cite{summer01data}. The present 95\% C.L.\ lower
bound on $\MH$ from the direct search at LEP of $\MH =
114.1$~GeV~\cite{LEPHiggsSM0701} is also indicated. The plot shows the
well-known preference for a light Higgs boson within the SM. Confronting
the theoretical prediction (allowing a variation of the top-quark mass,
$\mt$, which at present dominates the theoretical uncertainty, within
$1\sigma$) with the $1\sigma$ region of $\MW^{\mathrm{exp}}$ and the
95\% C.L.\ lower bound on $\MH$, one finds that at the present level of
accuracies the $1\sigma$ regions do no longer overlap.
 
\begin{figure}[htb]
\centerline{
\psfig{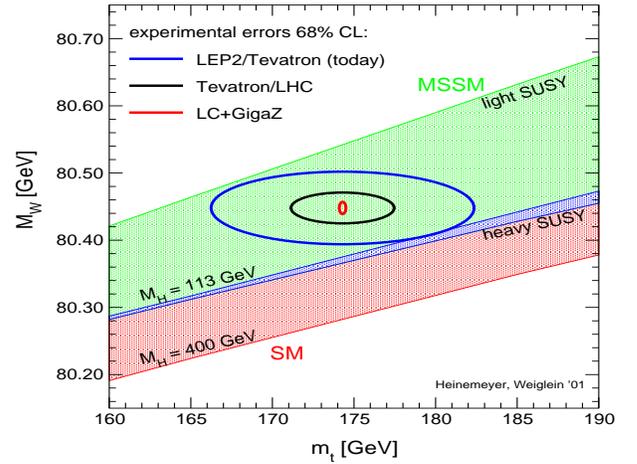}
}
\caption{The theoretical prediction for $\MW$ within the SM and the MSSM
in comparison with the current experimental accuracies (LEP2/Tevatron)
and the prospective accuracies at the LHC/LC and at GigaZ.
\label{fig:mwsmmssm}
}
\end{figure}

By comparing the SM predictions for the precision
observables with those of extended models, it can be investigated
whether the data allow a distinction between different kinds of
possible models. In \reffi{fig:mwsmmssm} the predictions for 
$\MW$ in the SM and the MSSM are shown as a function of $\mt$. The MSSM
prediction contains the dominant SUSY contributions of $\oaas$ and 
${\cal O}(\alpha^2)$ to the 
$\rho$~parameter~\cite{mssm2lqcd,mssm2lalpsq}. The allowed region in the SM
corresponds to varying $\MH$ in the interval $114 \gev \leq \MH \leq 400
\gev$, while in the region of the MSSM prediction the SUSY parameters
are varied, taking into account the constraints from direct searches for
SUSY particles. As indicated in the figure, the predictions in the SM
and the MSSM give rise to two bands with only a relatively small overlap
region. This region corresponds to the SM with a light Higgs, $\MH \lsim
130$~GeV, and to the MSSM with heavy superpartners, whose virtual
contributions decouple from the electroweak precision observables.

The predictions for $\MW$ in the SM and the MSSM are confronted in
\reffi{fig:mwsmmssm} with the current experimental accuracies of $\MW$
and $\mt$ (LEP2/Tevatron, outermost ellipse) and with the prospective
accuracies at the LHC and a future Linear Collider (LHC/LC) and at 
a high-luminosity Linear Collider running in a low-energy mode on the 
Z-boson resonance and the W-pair threshold (GigaZ). As can be read off
from the figure, the data on $\MW$ and $\mt$ presently show a slight
preference for the MSSM over the SM, which however statistically is not
very significant. The figure shows that the next generation of colliders,
in particular a Linear Collider in the GigaZ mode, promises an enormous
improvement in the experimental accuracies of $\MW$ and $\mt$ (and
furthermore also for $\sweff$) which will allow to test the
electroweak theory with unprecedented sensitivity~\cite{gigaz}.

Besides the indirect constraints from electroweak precision tests,
supersymmetric models provide a very
stringent direct test since they predict the existence of a relatively
light Higgs boson, whose mass can be calculated from the other
parameters of the model.

In \citere{mh2fd} a Feynman-diagrammatic result has been obtained 
for the dominant two-loop contributions to the masses of the neutral 
$\cp$-even Higgs bosons in the MSSM. The algebraic result obtained with
\fa\ and \two\ has been converted into \Fort\ code and has been
implemented into the program \fh~\cite{feynhiggs}.

\begin{figure}[htb]
\begin{picture}(220, 375)
\put(0,190){
\psfig{figure=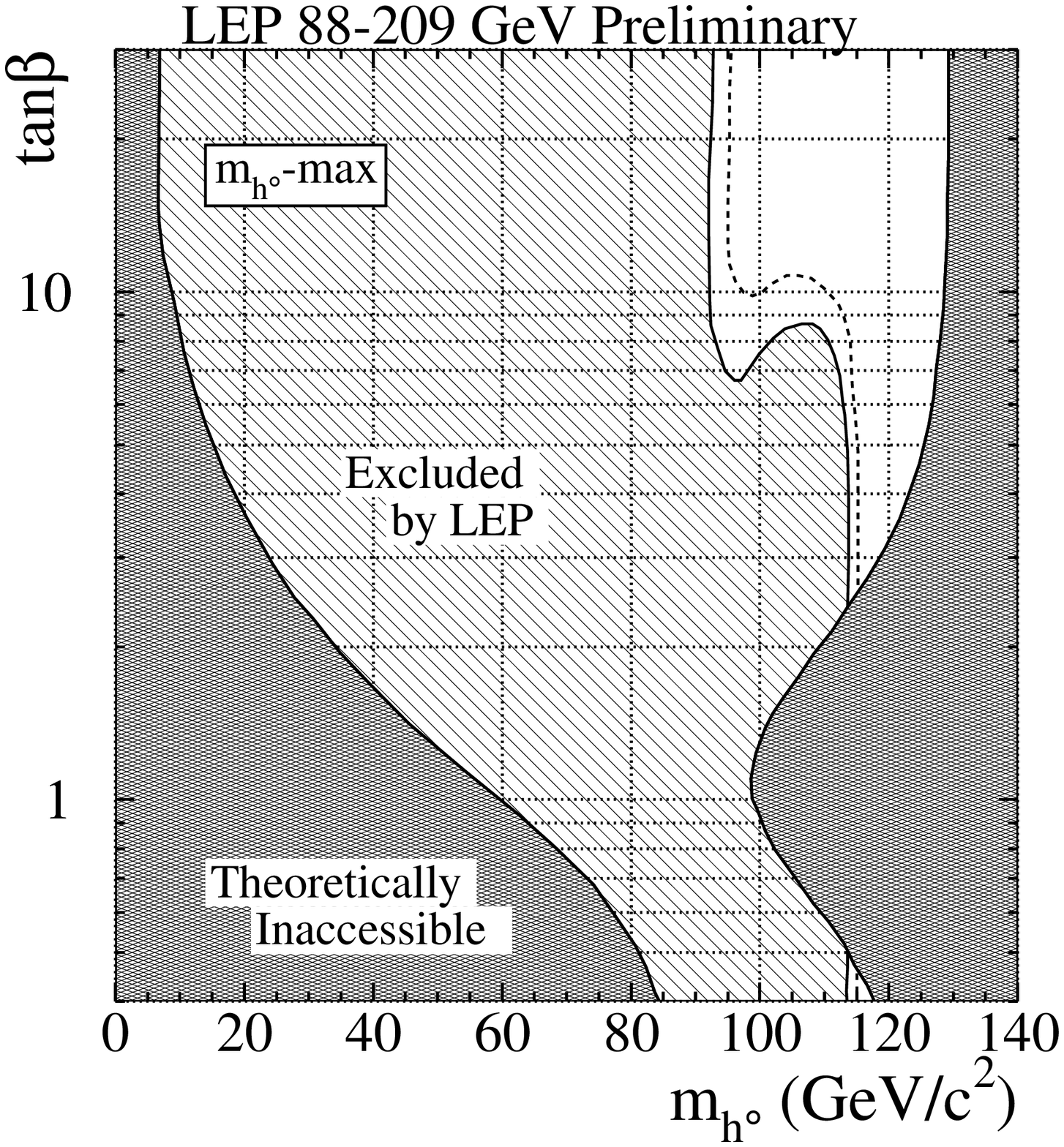,width=8cm,height=7.1cm}
}
\put(0,0){
\psfig{figure=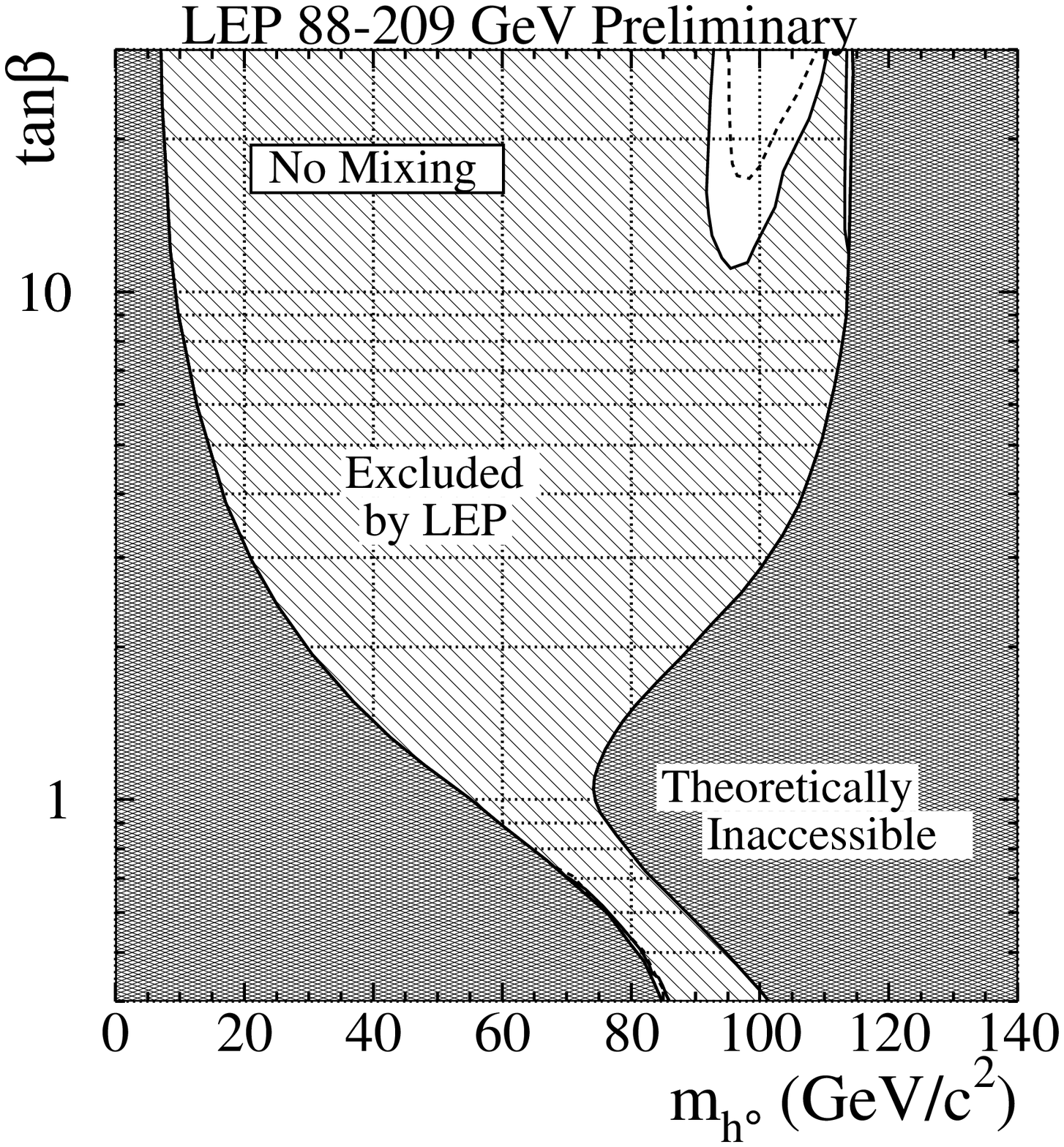,width=8cm,height=7.1cm}
}
\end{picture}

\caption{The 95\% C.L.\ bounds on $\mh$ in the $\mhmax$ and the
no-mixing benchmark scenarios obtained from combining the data of
the four LEP experiments are compared with the upper bound on $\mh$
within the MSSM~\cite{LEPHiggsMSSM}.
\label{fig:mhlep}
}
\end{figure}

While at the tree-level the lightest $\cp$-even Higgs boson in the MSSM
is bounded to be lighter than the Z-boson mass, this bound is shifted
upwards to $\mh \lsim 135$~GeV taking into account corrections up to the
two-loop order. The highest possible values for $\mh$ are obtained for
large values of $\tb$, the ratio of the vacuum expectation values of the
two Higgs doublets of the MSSM, large values of the mass of the $\cp$-odd 
Higgs boson, and a large mixing between the superpartners of the top quark.
Comparing the theoretical prediction for the upper bound on $\mh$ 
as a function of $\tb$ with the experimental exclusion limits obtained
at LEP2, it is possible to derive constraints on $\tb$. This is shown in
\reffi{fig:mhlep}, where the excluded region results from combining the
data of the four LEP experiments~\cite{LEPHiggsMSSM}, and the upper 
(and lower) bound within the MSSM (indicating the boundary to the 
``theoretically inaccessible'' region) has been obtained with \fh. 

The upper plot shows the case of the so-called ``$\mhmax$ benchmark 
scenario''~\cite{bench}, 
in which the MSSM parameters (for fixed values of $\mt = 174.3$~GeV and 
the SUSY scale $\msusy = 1$~TeV) are chosen such that $\mh$ as a function 
of $\tb$ takes its
maximal values. From the intersection of the experimentally excluded
region with the boundary to the theoretically inaccessible region one
finds an excluded region of $0.5 < \tan\be < 2.4$ within this scenario.
In the lower plot the MSSM parameters have been chosen according to the 
``no-mixing benchmark scenario''~\cite{bench}, which differs from the 
$\mhmax$ scenario in that vanishing mixing in the scalar top sector has been
assumed. In this case 
a much wider region of $\tb$ values, up to about $\tb \approx 10$, can be
excluded for $\mt = 174.3$~GeV and $\msusy = 1$~TeV.

\section{Conclusions and outlook}

The development of powerful computer-algebra systems was triggered by
applications in high-energy physics. Computer-algebra tools have
extensively been used in this field already for several decades, and
many of todays calculations would not have been feasible without computer 
algebra. A brief overview of computer-algebraic methods for the perturbative
evaluation of gauge theories has been given, and some examples have been
discussed of recent higher-order results obtained in the electroweak
Standard Model and its Minimal Supersymmetric extension.

The use of modern computer-algebra programs goes beyond their 
application as tools for certain steps of the calculations. As
indicated by the above examples of different collections of
programs for precision calculations within the
theory of the electroweak and strong interactions, an efficient 
communication of computer-algebra systems with other program components 
is of particular importance. These external programs can be packages 
for numerical evaluations, text processing tools, data bases, expert
systems, but also other computer-algebra programs being particularly
well suited for certain sub-parts of the problem. In order to facilitate
this kind of communication, the need for a certain degree of
standardization for the integration of program parts and the data
transfer between different systems will become more pronounced in the
future. 

Accordingly, future improvements of soft- and hardware promise a
further extension of the applicability of computer-algebra systems in
two different ways. On the one hand they could allow highly
sophisticated calculations which go beyond the scope of present
capabilities. On the other hand computer-algebra systems could more and
more become parts of general problem-solving environments, where
different components are integrated in such a way that the different
parts of calculations and the tasks of text processing, graphical
representation etc.\ are handled in the most efficient way.

\section{Acknowledgments}
The author thanks P.~Bhat, the other organizers of ACAT 2000, and the 
Fermilab Theory Group for the invitation and their kind
hospitality during his stay at Fermilab.

\nocite{*}
\bibliographystyle{aipproc}
\bibliography{test}

\end{document}

%% file: defs.tex
\def\beq{\begin{equation}}
\def\eeq{\end{equation}}
\def\beqar{\begin{eqnarray}}
\def\eeqar{\end{eqnarray}}
\def\barr#1{\begin{array}{#1}}
\def\earr{\end{array}}
\def\bfi{\begin{figure}}
\def\efi{\end{figure}}
\def\btab{\begin{table}}
\def\etab{\end{table}}
\def\bce{\begin{center}}
\def\ece{\end{center}}

\def\text{\textstyle}

\def\be{\beta}

\def\reffi#1{\mbox{Fig.~\ref{#1}}}

\def\citere#1{\mbox{Ref.~\cite{#1}}}

\newcommand{\ord}{{\cal O}}

\def\mathswitchr#1{\relax\ifmmode{\mathrm{#1}}\else$\mathrm{#1}$\fi}

\newcommand{\PW}{\mathswitchr W}

\newcommand{\PH}{\mathswitchr H}

\newcommand{\Ph}{\mathswitchr h}

\newcommand{\Pt}{\mathswitchr t}

\def\mathswitch#1{\relax\ifmmode#1\else$#1$\fi}

\newcommand{\MW}{\mathswitch {M_\PW}}

\newcommand{\MH}{\mathswitch {M_\PH}}

\newcommand{\Mt}{\mathswitch {m_\Pt}}

\newcommand{\mh}{\mathswitch {m_\Ph}}

\newcommand{\sweff}{\sin^2 \theta_{\mathrm{eff}}}

\def\tb{\tan\beta}

\newcommand{\mt}{\Mt}

\newcommand{\tsf}{\theta\kern-.20em_{\tilde{f}}}
\newcommand{\tsfp}{\theta\kern-.20em_{\tilde{f}\prime}}
\newcommand{\tsq}{\theta\kern-.15em_{\tilde{q}}}

\newcommand{\msusy}{M_{\mathrm{SUSY}}}

\newcommand{\lsim}
{\;\raisebox{-.3em}{$\stackrel{\displaystyle <}{\sim}$}\;}

\newcommand{\alps}{\alpha_{\mathrm s}}

\newcommand{\mma}{\textsl{Mathematica}}
\newcommand{\map}{\textsl{Maple}}
\newcommand{\FO}{\textsl{FORM}}
\newcommand{\Fort}{\textsl{Fortran}}
\newcommand{\fa}{\textsl{FeynArts}}
\newcommand{\fec}{\textsl{FeynCalc}}
\newcommand{\two}{\textsl{TwoCalc}}
\newcommand{\foc}{\textsl{FormCalc}}
\newcommand{\fh}{\textsl{FeynHiggs}}
\newcommand{\xlo}{\textsl{xloops}}

\newcommand{\oaas}{{\cal O}(\alpha\alps)}
\newcommand{\cp}{{\cal CP}}

\newcommand{\mhmax}{\mh^{\mathrm{max}}}

\newcommand{\VL}{\left( \begin{array}{c}}
\newcommand{\VR}{\end{array} \right)}
\newcommand{\ML}{\left( \begin{array}{cc}}
\newcommand{\MLd}{\left( \begin{array}{ccc}}
\newcommand{\MLv}{\left( \begin{array}{cccc}}
\newcommand{\MR}{\end{array} \right)}

\newcommand{\gev}{\,\, \mathrm{GeV}}

\newcommand{\BC}{\begin{center}}
\newcommand{\EC}{\end{center}}
\newcommand{\BE}{\begin{equation}}
\newcommand{\EE}{\end{equation}}
\newcommand{\BEA}{\begin{eqnarray}}
\newcommand{\BEAnn}{\begin{eqnarray*}}
\newcommand{\EEA}{\end{eqnarray}}
\newcommand{\EEAnn}{\end{eqnarray*}}

\newcommand{\id}{{\rm 1\kern-.12em
\rule{0.3pt}{1.5ex}\raisebox{0.0ex}{\rule{0.1em}{0.3pt}}}}

\def\draftdate{\relax}
\def\mda{\relax}
\def\mua{\relax}
\def\mla{\relax}
\def\draft{
\def\thtystars{******************************}
\def\sixtystars{\thtystars\thtystars}
\typeout{}
\typeout{\sixtystars**}
\typeout{* Draft mode!
         For final version remove \protect\draft\space in source file
*}
\typeout{\sixtystars**}
\typeout{}
\def\draftdate{\today}
\def\mua{\marginpar[\boldmath\hfil$\uparrow$]%
                   {\boldmath$\uparrow$\hfil}%
                    \typeout{marginpar: $\uparrow$}\ignorespaces}
\def\mda{\marginpar[\boldmath\hfil$\downarrow$]%
                   {\boldmath$\downarrow$\hfil}%
                    \typeout{marginpar: $\downarrow$}\ignorespaces}
\def\mla{\marginpar[\boldmath\hfil$\rightarrow$]%
                   {\boldmath$\leftarrow $\hfil}%
                    \typeout{marginpar:
$\leftrightarrow$}\ignorespaces}
\def\Mua{\marginpar[\boldmath\hfil$\Uparrow$]%
                   {\boldmath$\Uparrow$\hfil}%
                    \typeout{marginpar: $\Uparrow$}\ignorespaces}
\def\Mda{\marginpar[\boldmath\hfil$\Downarrow$]%
                   {\boldmath$\Downarrow$\hfil}%
                    \typeout{marginpar: $\Downarrow$}\ignorespaces}
\def\Mla{\marginpar[\boldmath\hfil$\Rightarrow$]%
                   {\boldmath$\Leftarrow $\hfil}%
                    \typeout{marginpar:
$\Leftrightarrow$}\ignorespaces}
\overfullrule 5pt
\oddsidemargin -15mm
\marginparwidth 29mm
}